\documentclass[12pt]{iopart}
\usepackage[dvips]{graphicx}% Include figure files

\begin{document}

\title{Quantifying Properties of the QCD Matter at RHIC}

\author{Huan Zhong Huang}

\address{Department of Physics and Astronomy, University of California, Los Angeles, CA 90095-1547, USA}

\address{Department of Engineering Physics, Tsinghua University, Beijing, 100084, China}

\address{\dag\ Email: huang@physics.ucla.edu}

\begin{abstract}

We will review recent results on quantitative description of global properties of bulk partonic matter at RHIC. These results include strangeness phase space factor of the partonic matter, azimuthal angular anisotropy $v_2$, and transverse momentum $p_T$ distributions of effective partons at the hadronization of bulk partonic matter. 
We present empirical constraints on parton energy loss in the high $p_T$ region ($>$ 5 GeV/c). A flat $R_{AA}$ as a function of $p_T$ at mid-rapidity implies a constant fraction of the parton energy loss ($\Delta p_T/p_T$) and the fraction reaches $25\%$ for neutral $\pi$, charged hadrons and non-photonic electrons of heavy quark decays from central Au+Au collisions at $\sqrt{s_{NN}}$ 200 GeV. Collision centrality dependence of $\Delta p_T/p_T$ from Au+Au and Cu+Cu collisions indicates that the fraction is approximately proportional to particle rapidity density $dn/dy$ divided by the initial transverse overlapping area of the colliding nuclei. Implications on dynamics of parton energy loss will be discussed.   

\end{abstract}

\section{Introduction}
High temperature and high energy density matter which exhibits partonic degrees of freedom has been produced in central nucleus-nucleus collisions at RHIC~\cite{star-white,phenix-white}. Two distinct experimental achievements at RHIC include studies on parton energy loss in the dense QCD medium and jet-medium interactions, and the establishment of collective bulk properties such as elliptic flow $v_2$ and constituent quark number scaling at the intermediate $p_T$ region. One prominent feature of the bulk partonic matter from high energy nucleus-nucleus collisions has been the unique hadronization scheme: the coalescence/recombination of effective partons to form hadrons~\cite{voloshin,ko,hwa,fries}. The constituent quark number scaling observed in $v_2$ and the baryon-meson grouping in the nuclear modification factor $R_{AA}$/$R_{CP}$ are consequences of the unique hadronization process for partonic matter.

The constituent entity of the partonic matter near the quark-hadron transition $T_c$ has not been determined. Theoretical considerations for a quasi-hadron system~\cite{brown-z} have been partially successful in explaining many experimental features. The nature of the constituents of the matter must also be time-dependence since the gluon degrees of freedom must dominate at mid-rapidity when two nuclei collide with each other. This conclusion is based on the quark and gluon distribution functions of the nucleus and the gluon-gluon scattering cross sections. The gluon degrees of freedom must disappear at hadronization. We do not know the evolution dynamics nor the exact nature of the entity at the hadronization. We use the term effective partons or effective constituent quarks to denote the empirical fact that the number of constituent quarks $n$ in a hadron seems to be an important factor in the hadron formation dynamics near $T_c$. 

The coalescence/recombination process may be applicable to hadronization at high parton density in general. Hwa and Yang applied the recombination model to d+Au collisions and their calculation matches experimental d+Au data~\cite{Hwa-dAu}. Preliminary STAR measurements of $R_{CP}$ for strange hadrons including multi-strange hyperons from d+Au collisions also show a baryon-meson grouping~\cite{haijiang}, a feature of coalescence/recombination mechanism. We use the coalescence/recombination framework to investigate effective quark properties from bulk matter at the time of hadronization.

\section{Global Properties of Effective Quark Degrees of Freedom}

\begin{figure}
\begin{center}
\includegraphics[width=0.40\linewidth,height = 4.0 cm,clip]{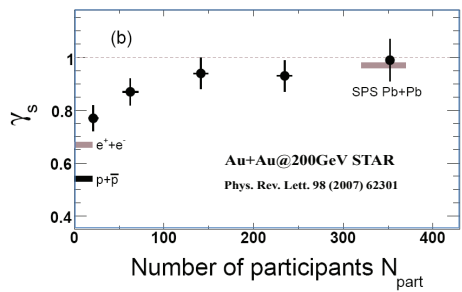}
\caption{Strangeness phase space occupancy factor $\gamma_s$ as a function of collision $N_{part}$.}
\label{fig1:Fig1}
\end{center}
\end{figure}

One of the global properties for bulk partonic matter is the strangeness content since strange quarks are produced during the collisions while up and down quarks can be produced or primordial from colliding nuclei. Figure~\ref{fig1:Fig1} shows the strangeness phase space occupancy factor $\gamma_s$ as a function of number of participant nucleons ($N_{part}$). The strangeness enhancement is one of the signatures for the Quark-Gluon Plasma (QGP) formation~\cite{JR-BM}. Gluon fusion process is an effective channel for strange quark pair production and the pair production rate is considerable when the temperature of the QGP is similar to or higher than the strange quark mass. Experimental data indicate that in the most central Au+Au collisions the $\gamma_s$ factor approaches one, an indication for strangeness equilibration in these collisions.

Thermal statistical models have been used to describe particle productions in nuclear collisions and the $\gamma_s$ factor, a parameter in the thermal statistical fit for particle yields, is model dependent. We have adopted a thermal statistical fit procedure, as described in reference~\cite{scaling,thermal}, where hadron yields are determined by a degeneracy factor and a statistical(Bose-Einstein or Fermi-Dirac) distribution. The strange quark phase-space occupancy factor $\gamma_s$ was introduced to describe possible lack of availability for strange quarks at the time of hadron formation. Generally in hadronic reactions strange hadrons are suppressed beyond their statistical and degeneracy factors ($\gamma_s < 1$). In this thermal statistical approach, the chemical freeze-out temperature parameter for RHIC data is approximately 165-170 MeV, close to the Lattice QCD prediction for critical temperature. The thermal parameters from the model may reflect the state of matter near the phase boundary between partons and hadrons. Hadron masses have been used to determine the statistical factor in thermal statistical models. However, it is likely that hadronic processes cannot maintain equilibrium states among vastly different hadrons during the rapid quark-hadron transition. Processes involving multiple entities have been proposed for the dynamics to equilibrium~\cite{PBM-thermal}. The success of thermal statistical description may reflect the quasi-hadron properties of the partonic matter prior to hadronization and the hadronization process would be a smooth collapse of the correlated quasi-hadron states to physical hadrons. In this picture our effective constituent quarks at the hadronization would correspond to the average of degrees of freedom in the quasi-hadron states.

Non-equilibrium thermal statistical model has also been used successfully to fit hadron yields from nucleus-nucleus collisions at RHIC~\cite{GT}. In this approach the temperature parameter is about $140$ MeV. The measured yields of baryons and hyperons are much higher than the statistical abundance allowed by such a relatively low temperature. Therefore, over-saturation of both up/down and strange quarks ($\gamma_q>1$ and $\gamma_s>1$) has to be introduced to accommodate the experimental data. Both equilibrium and non-equilibrium approaches have been able to match the measured hadron yields in the final state reasonably well. However, conceptually hadronic resonances would defy a sharp definition of chemical freeze-out temperature because these resonances continue to evolve during the hadronic stage. Perhaps non-equilibrium thermal statistical approach, due to the lower temperature parameter, may be able to account for some resonance evolution effect. 

\begin{figure}
\begin{center}
\includegraphics[width=0.45\linewidth,height = 5 cm,clip]{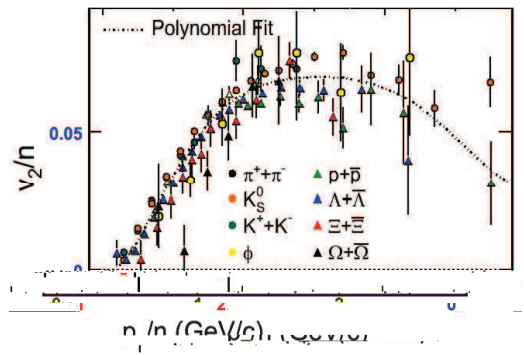}
\caption{$v_2/n$ as a function of $p_T/n$ for identified hadrons from Au+Au collisions. In the framework of coalescence/recombination models $v_2/n$ represents the azimuthal angular anisotropy developed by the effective contituent quarks.}
\label{fig2:Fig2}
\end{center}
\end{figure}

In the framework of coalescence/recombination model the effective quark degrees of freedom coalesce together to form a hadron. Figure~\ref{fig2:Fig2} shows elliptic flow $v_2/n$ as a function of $p_T/n$, where n is the number of constituent quarks in a hadron ($n=2$ for mesons and $n=3$ for baryons). This gross feature represents the azimuthal angular anisotropy of effective constituent quarks at hadronization. In the low $p_T$ region there is a hadron mass dependence and the quark number scaling is not exact. This is an indication that the effective constituent quark degrees of freedom depend on quark flavor and the effective mass. Hydrodynamic calculations in the low $p_T$ region provided a mass dependence of the $v_2$ values consistent with experimental observations. An important feature in the quark number scaling is that for $p_T/n > 1$ GeV/c the baryon $v_2/3$ and meson $v_2/2$ values fall on the same curve. In particular, the elliptic flow of $\phi$ meson, whose mass is close to that of baryon, seems to follow the meson scaling of $v_2/n$~\cite{star-phi}. Coalescence/recombination models for hadron formation provide natural explanation for the quark number scaling. Other hadronization mechanisms such as string fragmentation models would have very different dynamics for baryon versus meson formations, too complicated to yield such a simple scaling. 

It has been shown that the variable $K_{ET} = (m_T-m_0)/n$ appears to describe better the scaling of $v_2/n$ for variety of hadrons~\cite{phenix-prl,star-v2}, where $m_T=\sqrt{p_T^2+m_0^2}$ is the transverse mass and $m_0$ is the rest mass of hadrons. This is particularly interesting in the low $p_T$ region where there is a hadron mass dependence of elliptic flow $v_2$ as predicted by hydrodynamic calculations. Perhaps the variable $K_{ET}$, where hadron masses have been used, can capture better the mass and flavor dependence of effective constituent quarks in the quasi-hadron picture. The coalescence/recombination picture seems to apply in the low $p_T$ region as long as we take into account the collective flow developed during the partonic evolution by these effective constituent quarks prior to hadronization.

The $K_{ET}$ scaling seems to work surprisingly well for pions, which have major contributions from decay of resonances. The $v_2$ values as a function of $p_T$ often lie above the hydrodynamic calculations because of the resonance decay effect. Resonances suffer considerable perturbations during the hadronic evolution. We have advocated using $\Omega$, $\Xi$ and $\phi$ particles to investigate matter properties near the quark-hadron phase boundary. These particles are believed to have small hadronic re-scattering cross sections and, therefore, carry the information near the phase boundary. Recently we have developed a procedure to use transverse momentum spectra of these particles to extract the $p_T$ distribution of effective strange and up/down quarks in the bulk matter at hadronization.

Based on quark coalescence/recombination picture we assumed that baryons of momentum $p_T$ are mainly formed from quarks of momenta $\sim p_T/3$ on average, whereas mesons at $p_T$ are mostly from partons of $p_T/2$. The production probability of a baryon or a meson is proportional to the product of local parton density for the effective constituent quarks. We further assumed that the strange and anti-strange quarks have the same $p_T$ distribution. We argue that the effective strange and up/down quark
$p_T$ distributions can be extracted from $ s \sim \frac{\Omega(p_T/3)}{\phi(p_T/2)} $ and 
$ u/d \sim \frac{\Xi(p_T/3)}{\phi(p_T/2)} $.

\begin{figure}
\begin{center}
\includegraphics[width=0.45\linewidth,height = 5.0 cm,clip]{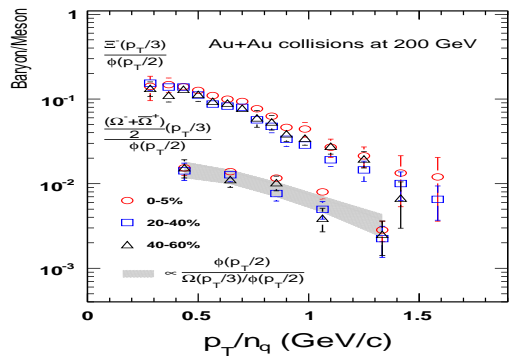}
\caption{Effective strange and up/down quark $p_T$ distributions empirically obtained in the framework of coalescence/recombination scheme.}
\label{fig3:Fig3}
\end{center}
\end{figure}

Figure~\ref{fig3:Fig3} shows s and u/d quark distributions as a function of the quark transverse momentum ($p_T/n$). These distributions represent the shapes of effective quark $p_T$ distribution in the bulk partonic matter at hadronization. We note that our approach is proven to be self-consistent empirically: the $\phi(p_T/2)/s$ (shaded area) ratio overlaps with the s quark distribution. Details of the analysis can be found in references~\cite{jhchen-prc}. 

We have relied on coalescence/recombination framework to obtain the quantitative description of azimuthal angular anisotropy $v_2$ and $p_T$ distributions for effective partons from the bulk QCD medium at hadronization. 
Hwa and Yang predicted that if $\Omega$ hyperons and $\phi$ mesons are solely from coalescence/recombination of thermal strange quarks, there will be no jet-like di-hadron correlations associated with $\Omega$ and $\phi$ particles~\cite{hwa-omega}. STAR reported preliminary results on two-hadron correlations in $\Delta \eta-\Delta \phi$ space which indicate that the jet-like two hadron correlation structure persists even to a $p_T$ range below a few GeV/c~\cite{trainor}. The interpretation for and implications of the correlation structure have not been established yet. The existence of the correlated structure does not necessarily render the coalescence/recombination framework invalid. It is an indication that the bulk partonic matter is not static: the matter may be perturbed by mini-jet or jet-like partons inducing kinematic correlations among partons in presence which may survive the hadronization process. The effective partons that coalesce/recombine to form hadrons are not exclusively from thermal partons. One or more constituent partons of hadrons in these correlations may be from mini-jet or jet-like partons. Our partonic $v_2$ and $p_T$ distributions using the coalescence/recombination framework are an average behavior of all partons: in the soft $p_T$ region thermal partons and mini-jet partons cannot be differentiated experimentally even though their momentum and correlation distributions may be different.   

\section{Empirical Constraints on Parton Energy Loss in QCD Dense Medium}

Measurements of nuclear modification factors, $R_{AA}$ and $R_{CP}$, have established that partons suffer considerable energy loss while traversing the high temperature and high energy density matter created in central nucleus-nucleus collisions at RHIC~\cite{star-white,phenix-white}. Critical aspects of dynamical description of parton energy loss in QCD medium include time-dependent evolution of the QCD dense medium, partonic energy loss mechanisms and geometry of colliding nuclei. The dynamical evolution of the QCD dense medium, especially the rapid change of local energy density, impacts parton energy loss processes. Many theoretical calculations of parton energy loss used a static medium, which may not truly capture the consequences of full time evolution. Recent theoretical advances include parton energy loss in a dynamical medium evolving according to hydrodynamic equations~\cite{renk,bass}. 

The mechanism for parton energy loss has been a subject of intensive theoretical investigations. The gluon radiative process has been considered a dominant mechanism for parton energy loss in the QCD medium~\cite{gyulassy}. Recent measurements of strong suppression of high $p_T$ non-photonic electrons from heavy quark decays are not consistent with the expectation of reduced energy loss for heavy quarks due to 'dead-cone' effect~\cite{dead-cone}. The collisional energy loss mechanism has been re-examined and has been found to be non-negligible in the $p_T$ region relevant for experimental measurements~\cite{collision}. One major uncertainty that severely hindered the interpretation of the measured nuclear modification factor for non-photonic electrons is the relative contribution to non-photonic electrons from Charm and Bottom meson decays. Experimentally electron-hadron correlations have been used to extract the relative B and C semi-leptonic decay contributions to non-photonic electrons~\cite{xlin}.

The geometry of the colliding nuclei is intrinsically convoluted with the dynamical processes of parton energy loss. In a parton energy loss calculation for a static medium, K. J. Eskola et al~\cite{eskola} pointed out that most surviving high $p_T$ particles are from the surface region of the participant volume of the colliding nuclei. Therefore, the core region of the dense medium can be practically opaque to the measured high $p_T$ particles and the inclusive single particle probe is fragile. Di-hadron correlations have been predicted to be more penetrating into the core region of the participant volume~\cite{zhang}. 

Empirically the nuclear modification factors ($R_{AA}$) for neutral pions, charged hadrons and non-photonic electrons are found to be approximately flat as a function of $p_T$ for $p_T$ greater than 5 GeV/c. For a collision centrality we treat the high $p_T$ spectrum from nucleus-nucleus collisions normalized by the number of binary collisions as equivalent to the high $p_T$ spectrum from p+p collisions with an energy loss of $\Delta p_T$. For a power law shaped high $p_T$ spectrum from p+p collisions the effective parton energy loss $\Delta p_T$ is approximately proportional to $p_T$ in order to obtain a flat $R_{AA}$ as a function of $p_T$. We examined the fraction of parton energy loss $\Delta p_T/p_T$ for various collision centralities of Au+Au and Cu+Cu collisions at $\sqrt{s_{_{NN}}}$ 200 GeV.

\begin{figure}
\begin{center}
\includegraphics[width=0.45\linewidth,height = 4.2 cm,clip]{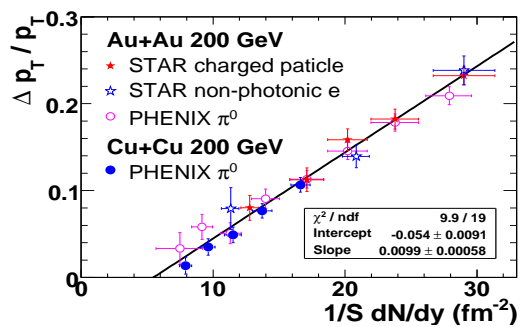}
\caption{Fraction of high $p_T$ particle energy loss as a function of $\frac{1}{S}\frac{dn}{dy}$, which can describe the unique
centrality dependence for both Au+Au and Cu+Cu collisions.}
\label{fig4:Fig4}
\end{center}
\end{figure}

We found that the quantity, $\frac{1}{S}\frac{dn}{dy}$, can provide unique characterization for the fraction of parton energy loss as a function of collision centrality for both Au+Au and Cu+Cu collisions, where $S$ is the initial transverse area of the overlapping nuclei and $\frac{dn}{dy}$ is the final particle rapidity density. Figure~\ref{fig4:Fig4} shows the fraction of
parton energy loss $\Delta p_T/p_T$ as a function of $\frac{1}{S}\frac{dn}{dy}$ for neutral pions, charged hadrons and non-photonic electrons from Au+Au and Cu+Cu collisions. Details of the analysis can be found in reference~\cite{gwang}. 

We note that the quantity $\frac{1}{S}\frac{dn}{dy}$ may be proportional to the initial local parton density. The experimental data seem to imply that there is no explicit path length dependence in $\Delta p_T/p_T$ for various collision centralities. We speculate that the absence of strong path length dependence may be due to the rapid decrease of local parton density and the energy loss of the traversing parton is a strong function of the local parton density of the medium. If the high $p_T$ parton loses its energy mostly in the first a few fm/c period, then the actual geometrical path length that the parton traverses in a static medium of participant matter will be irrelevant. As a consequence high $p_T$ partons which are originated at the core of the participant matter may have a finite probability to escape after suffering energy loss over a few fm/c time interval. 

Our considerations on the dynamical picture of parton energy loss have been based on gross features of centrality dependence of parton energy loss. The precise data from Cu+Cu collisions in the intermediate to low $N_{part}$ region provide important constraints to the trend. Recent theoretical calculations of parton energy loss use hydrodynamics to describe the evolution of the dense QCD matter, which may be able to capture the time evolution of local parton densities after the presumed equilibration time of 0.6-1.0 fm/c~\cite{bass,hirano}. We note that the non-equilibrium period before the hydrodynamics set in may also be important for parton energy loss because the local parton density may be very high during the non-equilibrium phase. 

The nuclear modification factor, $R_{AA}$ or $R_{CP}$, depends on dynamical convolution between parton energy loss mechanism and collision geometry. Di-hadron correlations have been studied as a function of reaction plane angle for triggered particles in Au+Au collisions at RHIC. There is an apparent path length effect for soft particles associated with the trigger particle around several GeV/c. Such a jet-medium interaction seems to depend on the path length that the jet traverses. Arguments have been made that these intermediate $p_T$ particles may contain contributions from non-jet processes~\cite{jia}, and possible bias from reaction plane determination in events with jet structure needs to be investigated also~\cite{gwang}.   

We argue that measurements of both $v_2$ and $R_{AA}$ for leading high $p_T$ particles will provide complementary constraints  that may allow us to differentiate parton energy loss mechanism from geometrical effects. The $p_T$ of the leading particle must be sufficiently high ($p_T > 5$ GeV/c) so that it will be in the regime dominated by dynamics of parton propagation and fragmentation. Dynamical pictures of parton energy loss, whether a complete opaque core of the participant matter in the fragile picture or a limiting energy loss in the first a few fm/c stage, will lead to different predictions for $v_2$ and $R_{AA}$ of leading high $p_T$ particles. Theoretical calculations on both $v_2$ and $R_{AA}$ for high $p_T$ leading particles will be needed to test dynamics of parton energy loss in the QCD medium.  

We thank Dr. Gang Wang, Dr. Jinhui Chen, Dr. Shingo Sakai, Prof. Yugang Ma and Prof. Charles Whitten for many stimulating discussions. This work is supported by a grant from U.S. DOE Office of Nuclear Physics and by National Natural Science Foundation of China.

\end{document}